\documentclass[aip,bmf, preprint]{revtex4-1}

\usepackage{graphicx}
\usepackage{multirow}
\usepackage[]{natbib}

\begin{document}

\title{Magnetic domain structure of La$_{0.7}$Sr$_{0.3}$MnO$_{3}$ thin-films probed at variable temperature with Scanning Electron Microscopy with Polarization Analysis }
\author{Robert M. Reeve}
\email[]{robert.reeve@dunelm.org.uk}
\author{Christian Mix}
\author{Markus K\"{o}nig}
\author{Michael Foerster}
\author{Gerhard Jakob}
\author{Mathias Kl\"{a}ui}
\affiliation{Institut f\"{u}r Physik, Johannes Gutenberg-University Mainz, 55099 Mainz, Germany.}

\begin{abstract}
The domain configuration of $50\;$nm thick La$_{0.7}$Sr$_{0.3}$MnO$_{3}$ films has been directly investigated using scanning electron microscopy with polarization analysis (SEMPA), with magnetic contrast obtained without the requirement for prior surface preparation. The large scale domain structure reflects a primarily four-fold anisotropy, with a small uniaxial component, consistent with magneto-optic Kerr effect measurements. We also determine the domain transition profile and find it to be in agreement with previous estimates of the domain wall width in this material. The temperature dependence of the image contrast is investigated and compared to superconducting-quantum interference device magnetometry data. A faster decrease in the SEMPA contrast is revealed, which can be explained by the technique's extreme surface sensitivity, allowing us to selectively probe the surface spin polarization which due to the double exchange mechanism exhibits a distinctly different temperature dependence than the bulk magnetization.\end{abstract}

\maketitle

In the development of materials for spintronic devices, magnetic oxides have been receiving considerable attention due to their range of exhibited properties (multiferroic, superconducting, birefringent \emph{etc.}), tunability and high theoretically predicted and experimentally observed spin polarization~\cite{oxides}. One particularly promising material is the half-metallic perovskite manganite La$_{0.7}$Sr$_{0.3}$MnO$_3$ (LSMO), whose measured spin polarization has been confirmed as nearly 100$\;\%$~\cite{LSMOpol} and which furthermore exhibits special magnetotransport effects such as a large collosal magneto-resistance (MR) due to the double exchange mechanism~\cite{CMR}. Furthermore, it exhibits a high Curie temperature ($T_c$) for this type of complex magnetic perovskite oxide, around 350$\:$K~\cite{LSMOqual}, thereby enabling room temperature applications. In order to elucidate the key physics governing the magnetic properties of these films on the length scales of interest for future devices, a fundamental step is the imaging of the magnetic domain configuration of the films with high spatial resolution, since the domain configuration affects key parameters such as the MR~\cite{MR}. Previously the domain structure of LSMO films/elements has been investigated with Kerr microscopy~\cite{Kerr}, magnetic-force microscopy~\cite{MFM, MR, sputterprob} and X-ray magnetic microscopy~\cite{patLSMO, sputterprob, Taniuchi2, heidler}, however, these techniques are either limited in their resolution or require the use of large scale facilities. Scanning electron microscopy with polarization analysis (SEMPA) is one suitable technique which can be employed to directly view the magnetization configuration of a variety of materials and which permits the quantiative determination of the magnetization vector, with a typical resolution better than 20$\:$nm~\cite{SEMPArev}. Furthermore it allows one to probe the spin polarization of the material by measuring the contrast provided by the secondary electron spin polarization. Previous work has applied the technique to bulk crystals of LSMO and related oxides~\cite{LSMOSEMPA1, LSMOSEMPA2} but imaging of the technologically relevant thin films has been lacking. Additionally, for a full understanding it is necessary to image as a function of temperature, down to cryogenic conditions, since changes in the films properties have been seen~\cite{tempchange}. In particular, important MR effects such as tunneling MR (TMR) are experimentally observed to display attractive values at very low temperatures, but exhibit a significant reduction as the temperature is increased~\cite{TMR1, TMR2}. In this paper we are able to image the spin structure in epitaxial thin films of the material which are desirable for future applications and where the geometry and substrate can be used to tailor the magnetization configuration~\cite{Perna, MFM}.  Furthermore we determine the temperature dependence of the spin polarization of the secondary electrons and compare this with the temperature dependence of the magnetization. Through the surface sensitivity of the technique we show that we are able to selectively probe with a high spatial resolution the upper interface of the film, which can be expected to dominate the properties of interest for many technologically important applications and the results can in turn be related directly to the fundamental origin of the magnetic interactions in the material.
\\50$\:$nm LSMO thin films were deposited by pulsed laser deposition (KrF, \emph{Compex Pro}, 248$\:$nm, 20$\:$ns) in an ultra-high vacuum (UHV) chamber with a base pressure below 10$^{-7}\:$mbar. The target to substrate distance is fixed at 50$\:$mm. Optimum substrate temperature and deposition pressure have been found as 580$\:^\circ$C and 0.15$\:$mbar, respectively.  The laser energy and repetition rate were kept at 2.1-2.3$\:$J/cm$^2$ and 5$\:$Hz. Films were deposited on (001) TiO$_2$-terminated SrTiO$_3$ (STO) substrates, with a deposition rate of 1.4$\:$\AA/s. After deposition the samples were annealed at 550$\:^\circ$C and 100$\:$mbar oxygen pressure for one hour and cooled down slowly at the same pressure. To prevent water vapor adsorption the samples were rapidly transferred \emph{ex-situ} to our UHV SEMPA preparation and analysis system while still warm. The \emph{Omicron} SEMPA system consists of a UHV \emph{Gemini} scanning electron microscope in conjunction with a spin polarized low-energy electron diffraction (LEED) spin detector~\cite{SPLEED}, which can simultaneously measure both in-plane directions of the magnetization through the asymmetry in electron scattering into opposite LEED spots. The measurement stage incorporates a liquid He flow cryostat, permitting the cooling of samples down to below 30$\:$K. The base pressure of the chamber is below $3\times10^{-11}\:$mbar and the pressure remains in this regime during imaging. Frequent flash cleaning of the W(100) LEED crystal was performed between images in order to ensure that artificial reductions in image contrast from a deteriorating scattering surface are avoided.

\begin{figure}[tb!]

             \begin{center}

(a)            \includegraphics[width=0.9\columnwidth]{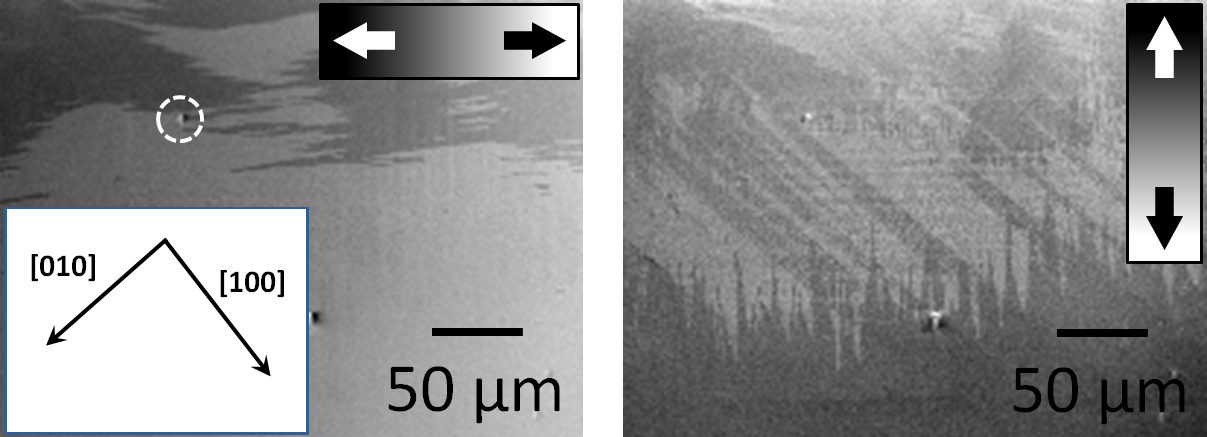}
\\(b)          \includegraphics[width=0.9\columnwidth]{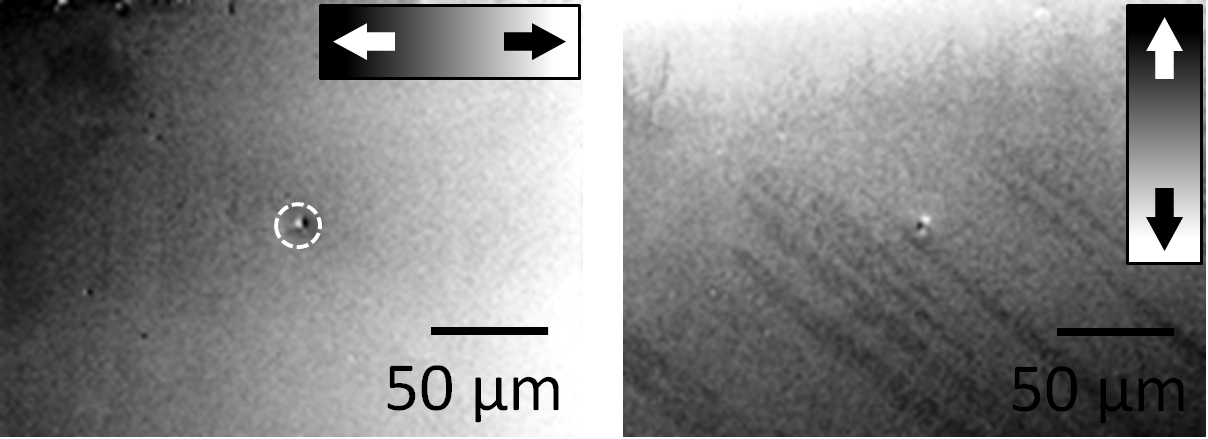}

           \end{center}

             \caption{Domain structure of LSMO thin films at 30$\;$K a) Large scale domain structure, indicating the formation of domains near the sample edge (top). The inset displays the crystallographic directions of the substrate. b) The change in domain structure following heating above $T_c$ and recooling. The scales on the images are qualitative guides to the magnetic contrast. The dotted area indicates a dirt particle which can be used to compare image location.}

             \label{SEMPApic}
\end{figure}

\begin{figure}[tb!]

             \begin{center}

(a)           \includegraphics[width=0.9\columnwidth]{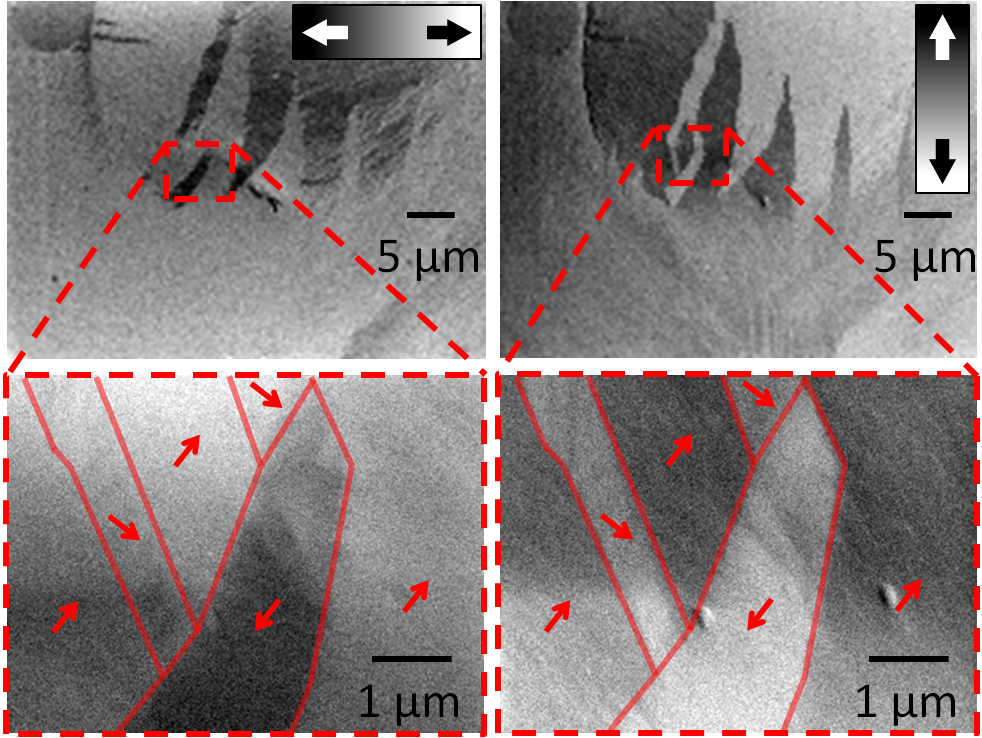}
\\(b)         \includegraphics[width=0.5\columnwidth]{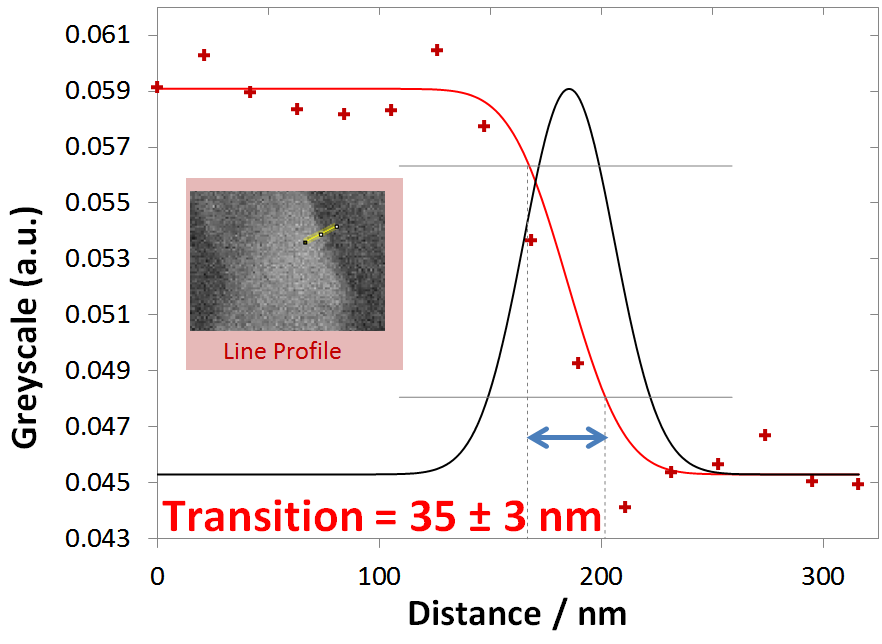}

           \end{center}

             \caption{High resolution images of the domain structure. The scales on the images are qualitative guides to the magnetic contrast. The arrows are schematic representations of the magnetization direction. The line scan in (b) indicates a domain transition of $35\pm3\;$nm. The substrate orientation is the same as in Fig. 1.}

             \label{SEMPApic2}
\end{figure}

\begin{figure}[t!]

             \begin{center}

             \includegraphics[width=1.0\columnwidth]{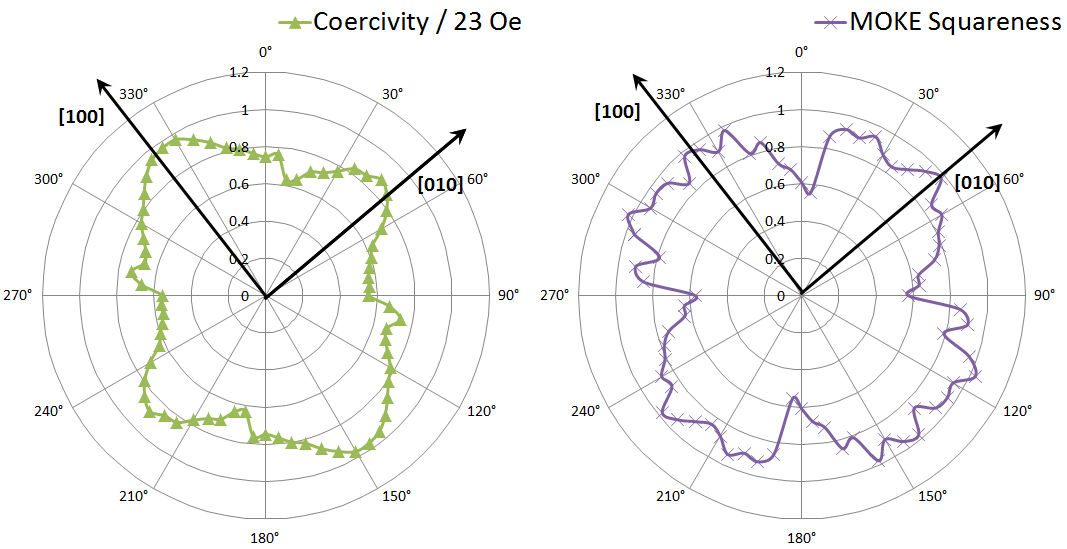}

           \end{center}

             \caption{The coercivity and squareness of the MOKE hysteresis loops as a function of sample angle. The data have been averaged for corresponding angles separated by 180$^\circ$.}

             \label{MOKE}
\end{figure}

Images of the domain structure of the films can be seen in Figure~\ref{SEMPApic}, where light and dark contrast indicate magnetization aligned parallel/antiparallel to the indicated in-plane film axis. The ability to image the film without prior treatment is surprising, since the extreme surface sensitivity of SEMPA  ($\sim$1$\:$nm) usually requires either \emph{in-situ} film growth or necessitates the cleaning of the samples before imaging. Such preparation procedures for clean surfaces can include \emph{in-situ} cleaving of bulk samples~\cite{LSMOSEMPA2}, dusting of the surface with iron~\cite{iron} or sputter cleaning of the film~\cite{sputter}. In the latter case, however, this can additionally lead to damage of the films and a change in the magnetic properties~\cite{sputterprob}. The fact that this preparation was not required for our films demonstrates their chemical stability and confirms that unwanted processes during growth, such as the emergence of Mn$_3$O$_4$ precipitates or Sr surface segregation as observed in some previous studies~\cite{LSMOqual}, are insignificant for our samples. This shows that in contrast to all materials that we have previously studied (3$d$ metals, Fe$_3$O$_4$, \emph{etc}.), contrast can be obtained on LSMO even without \emph{in-situ} cleaning  and with our probing depth of about 1 nm, we can exclude a larger contamination of the surface. The magnetic domain walls were observed to appear towards the edges of the film, which is a common feature of many continuous thin magnetic films since it is energetically favorable for multiple domains to form in this region in order to reduce the stray field,  while the bulk of high quality thin films is in a monodomain state. This can be seen in panel (a) where the domains are seen near the top of the image, which is adjacent to the sample edge, whilst near the bottom the films are in a continuous domain state. The left polarization component is mainly characterized by large features of tens of microns in size, while additional stripe like features are visible in the right component of the polarization. Rotating the detector changed the relative contribution of these different components to the two respective images (as the alignment of the magnetization and measurement axis varied), confirming that the features were magnetic in origin. The circled feature is a dirt particle which was used to aid positioning. The domain structure was reproducibly obtained on successive cycles of heating to ambient conditions and re-cooling, with the large scale features remaining unchanged throughout the temperature range. Following \emph{in-situ} heating to 450$\:$K, above $T_c$, the domain structure was observed to change. The new domain structure from the same region of the sample is displayed in panel (b). The magnetization is now largely featureless in this region, however, some striped features remain which bear a striking resemblance to those observed previously. By comparing the images to the crystallographic orientation of the substrate determined using X-ray diffraction, depicted in (a), it can be seen that the axis of the stripes corresponds to one of the principle axes, implying that the substrate morphology is influencing the domain structure, with step edges acting as pinning sites. High resolution images of the magnetization from elsewhere on the sample, after the change of domain structure, can be seen in Figure~\ref{SEMPApic2}. In (b) an integrated Gauss function has been fitted to a line scan across a domain wall, revealing a transition region of 35$\pm3\:$nm, which compares favourably with estimates of the domain wall width in this material ($\sim 30-55\:$nm~\cite{DWidth1, DWidth2}).
Furthermore, it can be inferred that a largely four-fold anisotropy exists in the films, as schematically indicated by the arrows in enlarged images of panel (a). To understand and corroborate the relationship between the observed domain structure and the inherent magnetic properties of the films, the hysteretic behavior of the films was probed using the magneto-optic Kerr effect (MOKE), at ambient conditions. The MOKE loop coercivity and squareness (the ratio of the signal at remanence to that at saturation) as a function of the angle of the sample with respect to the applied field are presented in Figure~\ref{MOKE}. 
The data confirm a primarily four-fold magnetic anisotropy, with the easy axes of the films aligned approximately with the in-plane $\langle100\rangle$ crystal axes. Additionally a small uniaxial contribution to the magnetic anisotropy can also be discerned from the symmetry of the coercivity plot. The anisotropy has previously been shown to be very dependent on the level of tensile or compressive strain in the films, which originates from the mismatch with the substrate~\cite{Perna, Steenbeck, Suzuki1, CMR}. A uniaxial component to the magnetic anisotropy has also been noted before and correlated to the presence of steps from the substrate, as seen with atomic force microscopy~\cite{Tsui, Taniuchi2, Perna}. Depending on the exact film substrate, thickness~\cite{Taniuchi2} or temperature~\cite{Tsui} either the biaxial or uniaxial anisotropy dominate the behaviour. In our 50$\;$nm film there appears to be a competition between these two components, in good agreement with Ref.~\cite{Taniuchi2} where a change from a uniaxial domain structure to a fourfold one was imaged as the film thickness was increased from 20 to 120$\;$nm. Overall the room temperature MOKE data show good agreement with the low temperature LSMO SEMPA imaging, supporting both a dominant four-fold anisotropy and an influence of the substrate on the domain pattern.

\begin{figure}[tb!]

             \begin{center}

             \includegraphics[width=0.9\columnwidth]{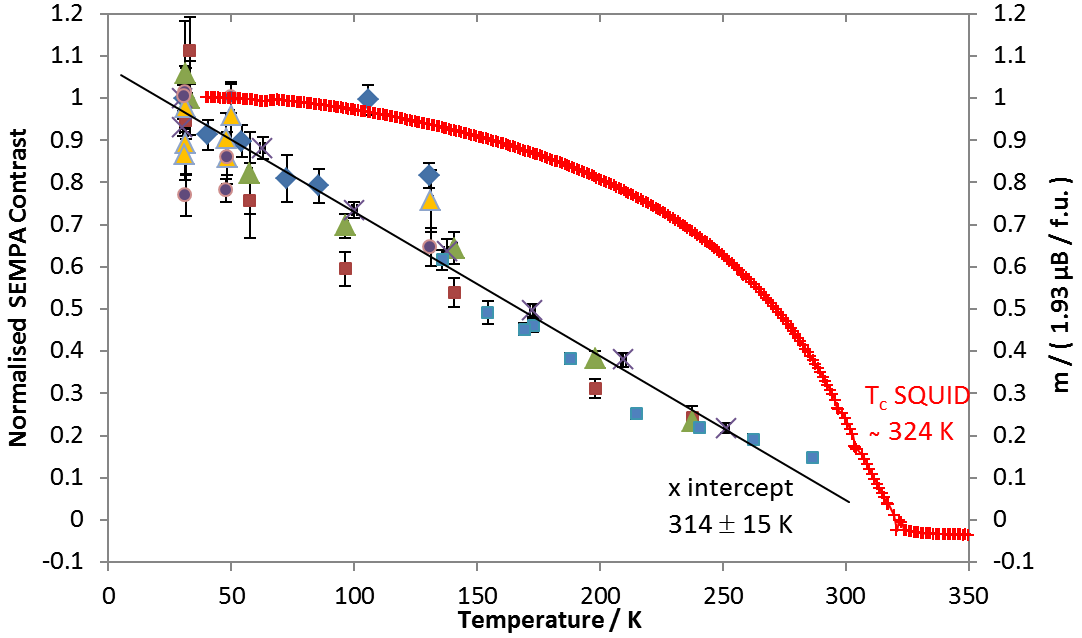}

           \end{center}

             \caption{Temperature dependence of the SEMPA image contrast as compared to the sample magnetization from SQUID at 500$\;$Oe applied field. Different symbols differentiate data from different experimental runs and polarization components.}

             \label{temp}
\end{figure}

Finally our setup uniquely allows us to study the surface spin polarization as a function of temperature. Figure~\ref{temp} displays the temperature dependence of the SEMPA image contrast in comparison with the magnetization curve obtained using superconducting quantum interference device (SQUID) magnetometry. The SQUID data show the expected decay of the magnetization with temperature, revealing a $T_c$ of 324$\:$K and a magnetic moment of about 1.93$\:\mu_{B}$/f.u.$\:$. This is in good agreement with the range of literature values for LSMO~\cite{LSMOqual}. The SEMPA contrast, meanwhile, can be described by a linear decrease as indicated by the regression line. Contrast is still directly discernable up to around 260$\:$K and by extrapolating the trend-line the contrast may be expected to completely vanish at 314$\pm$15$\:$K, in agreement with the $T_c$ obtained from the SQUID measurements. Nevertheless, the linear decrease in contrast clearly corresponds to a faster decay than the bulk magnetization.

The linear decrease must be directly related to the spin polarization of the secondary electrons originating from the film, since a change in the spin-dependent scattering at the detector crystal can be ruled out due to our excellent chamber pressure and frequent checks and flash cleaning of the scattering crystal~\cite{SPLEED}. The discrepancy can be explained by the extreme surface sensitivity of SEMPA imaging which means we probe only the surface magnetization of the film. Previous studies have revealed that the surface magnetism of perovskite manganites~\cite{temp2, temptheory} can exhibit a faster decay with temperature than bulk magnetization and in the case of LSMO a similar linear-like decrease has been observed in spin-resolved photoelectron spectroscopy data~\cite{temp1}. In such a material, the ferromagnetic alignment arises from the double-exchange mechanism, involving electron hopping between mixed valence manganese ions. The reduced symmetry of the surface restricts such hopping and this, in turn, can lead to a suppression in the surface magnetism~\cite{temp3} with a different temperature dependence than that of the bulk. In particular the 3$d$ electrons show this reduced spin polarization~\cite{temp1} and since it is the spin polarization of the 3$d$ electrons that in turn leads as a first approximation to the spin polarization of the charge carriers at the Fermi level, which dominate our signal, we can explain our findings by this mechanism. This shows that by SEMPA we can probe distinct spin polarization properties that do not necessarily reflect the bulk magnetization and thus deduce complementary information about the spin polarization in a spatially resolved manner.

In conclusion, we have demonstrated the ability to measure the domain structure of LSMO films using SEMPA in a technologically relevant thin film geometry, to our knowledge for the first time. Unusually the imaging did not necessitate prior surface preparation, implying a high quality and stability of the grown films, which is an important attribute in terms of the use of LSMO in future applications. The magnetic characterization measured a domain transition profile of around 35 nm and revealed a primarily four-fold anisotropy with easy axes along the $\langle$100$\rangle$ crystallographic directions. Additionally a small influence of substrate steps and terraces is suggested from both the SEMPA images and a uniaxial component of the anisotropy observed in MOKE measurements. Furthermore, the temperature dependence of the SEMPA images has been compared with the magnetization from SQUID data, revealing a faster decrease with temperature for the magnetization at the surface of the film. This result, which can be attributed to the suppression of electronic hopping at the surface, combined with the high spatial resolution of the images, highlights our ability to locally probe the surface-interface properties of the films which often govern properties of paramount importance for future devices based on spin transport across interfaces. In this regard, the rapid decay of the surface magnetization represents a potential limit to future device efficiencies in the vicinity of room temperature as it supports the theory that the similar rapid decay observed in MR signals arises from the fundamental properties of the surface~\cite{TMR1, TMR2}. The next step is to extend our imaging to patterned samples in order to further investigate the interplay between the magnetization and geometrical confinement in this material.

This work was funded by the EU's 7th Framework Programme IFOX (NMP3-LA-2012246102), MAGWIRE FP7-ICT-2009-5 257707), the European Research Council through the Starting Independent Researcher Grant MASPIC (ERC-2007-StG 208162), the Swiss National Science Foundation, and the DFG.

\bibliography{paper2}

\end{document}